\documentstyle[12pt]{article}

= -0.5 cm

\newcommand{\e}{{\rm e}}

\newcommand{\ug}{\; = \;}

\newcommand{\crm}{{\rm c}}

\newcommand{\GHz}{{\rm GHz}}
\newcommand{\constant}{{\rm constant}}
\newcommand{\SiO}{{\rm SiO}}

\newcommand{\text}{\rm}

\newcommand{\drm}{{\rm d}}

\newcommand{\mrm}{{\rm m}}

\newcommand{\la}{\lambda}

\newcommand{\ze}{\zeta}

\newcommand{\bb}{\begin{equation}}
\newcommand{\ee}{\end{equation}}
\newcommand{\bega}{\begin{eqnarray}}
\newcommand{\ega}{\end{eqnarray}}
\newcommand{\begae}{\begin{eqnarray*}}
\newcommand{\egae}{\end{eqnarray*}}

\newcommand{\h}{\hspace*{4ex}}
\newcommand{\dis}{\displaystyle}

\newcommand{\th}{\theta}

\newcommand{\om}{\omega}

\newcommand{\cent}{\centerline}
\newcommand{\vs}{\vspace*}

\begin{document}

\baselineskip 0.65cm

\begin{center}

{\large {\bf Localized Superluminal solutions to the wave equation in 
(vacuum or) dispersive media, for arbitrary frequencies and with adjustable
bandwidth}$^{\: (\dag)}$}
\footnotetext{$^{\: (\dag)}$
Work supported by FAPESP (Brazil), and by MIUR, INFN (Italy).
\ E-mail address for contacts: mzamboni@ifi.unicamp.br }

\end{center}

\vs{3mm}

\cent{ M. Zamboni-Rached, K. Z. N\'obrega, H. E.
Hern\'{a}ndez-Figueroa}

\centerline{{\em DMO--FEEC, State University at Campinas,
Campinas, S.P., Brazil.}}

\centerline{\rm and}

\cent{ Erasmo Recami }

\cent{{\em Facolt\`a di Ingegneria, Universit\`a statale di
Bergamo, Dalmine (BG), Italy;}}

\cent{{\em {\rm and} INFN--Sezione di Milano, Milan, Italy.}}

\vs{1.cm}

{\bf Abstract  \ --} \h In this paper we set forth new exact analytical
Superluminal localized
solutions to the wave equation for arbitrary frequencies and adjustable
bandwidth. The formulation presented here is rather simple, and its results
can be expressed in terms of the ordinary, so-called ``X-shaped waves". \
Moeover, by the present formalism we obtain the first {\em analytical}
localized Superluminal approximate solutions which represent beams
propagating {\em in dispersive media}. \ Our solutions may
find application in different fields, like optics, microwaves, radio waves,
and so on.

\vs{0.3cm}


\vs{0.3cm}


\newpage

{\large\bf 1. -- Introduction}\\

\h For many years it has been known that localized
(non-dispersive) solutions exist to the (homogeneous) wave
equation[1,2,3], endowed with subluminal or
Superluminal[4,5,6,7,8,9] velocities. These solutions propagate
without distortion for long distances in vacuum.

\h Particular attention has been paid to the localized
Superluminal solutions like the so-called X-waves[5,6,8] and
their finite energy generalizations[7,8]. It is well known that
such Superluminal Localized Solutions (SLS) have been {\em experimentally}
produced in acoustics[10], optics[11] and more recently microwave physics[12].

\h As is well known, the standard X-wave has a broad band
frequency spectrum, starting from zero[8,9] (it being therefore
appropriate for low frequency applications). This fact can be
viewed as a problem, because it is difficult or even impossible to
define a carrier frequency for that solution, as well as to use it
in high frequency applications.

\h Therefore, it would be very interesting to obtain exact SLSs to
the wave equations with spectra localized at higher (arbitrary)
frequencies and with an adjustable bandwidth (in other words, with
a well defined carrier frequency).

\h To the best of our knowledge, only two attempts were made in
this direction: one by Zamboni-Rached {\em et al}.[8], and the other by
Saari[13]. The former showed how to shift the spectrum to higher
frequencies without dealing with its bandwidth, while the latter
worked out an analytical approximation to optical pulses only.

\h In this work we are presenting analytical and exact
Superluminal localized solutions in vacuum, whose spectra can be
localized inside any range of frequency with adjustable bandwidths, and
therefore with the possibility of choosing a well defined carrier
frequency. In this way, we can get (without any approximation) radio,
microwave, optical, etc., localized Superluminal waves.

\h Taking advantage of our methodology, we obtain the first
analytical approximations to the SLS's in dispersive media (i.e.,
in media with a frequency dependent refractive index).

\h One of the interesting points of this work, let us stress,
is that all results are obtained from simple mathematical operations on
the standard ``X-wave".

\

\

{\large\bf 2. -- Superluminal localized waves in dispersionless media}\\

\h Let us start by dealing with SLSs in dispersionless media.
From the axially symmetric solution to the wave
equation in vacuum ($n=1$), in cylindrical
coordinates, one can easily find that

\begin{equation} \psi (\rho,z,t) \ug J_0(k_{\rho}\rho) \;
\dis{\e^{+ik_z z} \; \e^{-i\om t}}\label{eq1}
\end{equation}

with the conditions

\begin{equation} k_{\rho}^2 = \dis{\frac{\om^2}{c^2} - k_z^2} \; ; \ \
\ \ \ \ k_{\rho}^2 \geq 0 \ , \label{condicao}
\end{equation}

where $J_0$ is the zeroth-order ordinary Bessel function; $k_z$
and $k_{\rho}$ are the axial and the transverse wavenumber
respectively, $\om$ is the angular frequency and $c$ is the light
velocity.

\h It is essential to call attention right now to the dispersion
relation (\ref{condicao}). Positive (but not constant, a priori)
values of $k_{\rho}^2$, with real $k_z$, do allow {\em both} subluminal {\em and}
Superluminal solutions, while implying truly propagating waves
only (with exclusion of the evanescent ones). We shall pay
attention in this paper to the Superluminal solutions. Conditions
(\ref{condicao}) correspond in the ($\om,k_z$) plane to confining
ourselves to the sector shown in Fig.1; that is, to the region
delimited by the straight lines $\om = \pm ck_z$.

\h Important consequences can be inferred from (\ref{condicao}),
when performing the coordinate transformation

\begin{equation}
\left\{
\begin{array}{l}
\om=\om  \\
k_{z}=(\om/c) \, \cos \, \th \, ,
\end{array}
\right. \label{transform}
\end{equation}

which yields

\begin{equation}
k_{\rho}=(\om/c) \, \sin \, \th \, .
\label{Krho}
\end{equation}

 \h With this transformation, solution (\ref{eq1}) can be
rewritten, in the new coordinates ($\om$,$\th$), as

\begin{equation} \psi (\rho,\ze) \ug J_0(\frac{\om}{c} \rho \sin\th) \;
\dis{\e^{+i\om\ze \cos\th}}\,\,\,\, , \label{bessel}
\end{equation}

where $\ze\equiv z - Vt$, and where the propagation speed (group
velocity) is obviously $V=c/\cos\,\th$.  Equation (\ref{bessel})
states that the beam is transversally localized in energy, and propagates
without suffering any dispersion. It should be noticed also
the relationship between $V$ and $\th$: namely, each value of
$\th$ yields a different wave velocity. This fact will
be used in the next Section.

\h Equation (\ref{bessel}) represents the well known ``Bessel beam''. As can
be seen, such an equation has two free parameters,
$\om$ and $\th$. Considering $\th$ constant, and making a
superposition of waves for different frequencies, one can obtain
localized (non-dispersive), Superluminal solutions; namely

\begin{equation} \Psi (\rho,\ze) \ug
\int_{0}^{\infty}S(\om)J_0(\frac{\om}{c} \rho \sin\th) \;
\dis{\e^{+i\om\ze \cos\th}} \drm\om \; . \label{solgeral}
\end{equation}

\h In eq.(\ref{solgeral}), if an exponential spectrum\footnote{It is easy
to see that this spectrum starts from zero, it being suitable for
low frequency applications, and has the bandwidth $\Delta\om = 1/a$}
like $S(\om)=\e^{-a \om}$ is considered, one obtains,
by use of identity (6.611.1) of ref.[14],
the ordinary X-shaped wave:

\begin{equation} \Psi (\rho,\ze) \ug
\frac{1}{\sqrt{(aV-i\ze)^2+(\frac{V^2}{c^2}-1)\rho^2}},
\label{ondaX}
\end{equation}

where $a$ is a positive constant.

\h This solution is a wave that propagates in free space without
distortion and with the Superluminal velocity $V=1/\cos\,\th$.
Because of its non-dispersive properties, and its low frequency
spectrum, the X-wave is being particularly applied in fields like
acoustics[5]. The illustration of an X-wave, with parameters $a=
10^{-7} \;$s and $V=5 \, c$, is shown in Fig.2.

\

\

{\large\bf 3. -- Superluminal localized waves for arbitrary
frequencies and adjustable bandwidths}\\

\h In the last Section, it has been shown that a superposition of
Bessel beams can be used to obtain a localized and Superluminal
solution to the wave equation in a dispersionless medium. It is known
that it may be a difficult task, it being possible, or not, finding analytical
expressions for eq.(\ref{solgeral}). Its numerical solutions
usually brings in some inconveniences for further analysis,
uncertainties concerning the fast oscillating field components, etc.;
besides implying a loss in the physical interpretation of the
results. Thus, it is always worth looking for analytical
expressions.

\h Actually, the kind of solution found by us for eq.(\ref{solgeral})
is strictly related to the chosen spectrum $S(\om)$.
Following previous work of ours[8], we are going to present our
spectrum together with its main characteristics.

\

{\bf 3.1 -- The $S(\om)$ spectrum}\\

 \h One of our main objectives is finding out a spectrum which can preserve
the integrability of eq.(\ref{solgeral}) for any frequency range. In order
to be able to shift our spectrum towards the desired frequency, let us locate
it around a central frequency, $\om_\crm$, with an arbitrary bandwidth
$\Delta\om$.

\h Then, let us choose the spectrum

\begin{equation}
S(\om)=\left(\frac{\om}{V}\right)^m\e^{-a\om}
 \label{spectro}
\end{equation}

where $V$ is the wave velocity, while $m$ and $a$ are free
parameters. For $m=0$, it is $S(\om)=\exp[-a\om]$, and one
gets the (standard) X-wave spectrum.

\h After some mathematical manipulations, one can easily find the
following relations, valid for $m\neq 0$:

\

\hfill{$
\dis{m = \frac{1}{\left(\Delta\om_{\pm}/\om_\crm\right)-\ln\left(1+
\left(\Delta\om_{\pm}/\om_\crm\right)
\right)}}
$\hfill} (8.1)

\

\hfill{$
\dis{\om_\crm = \frac{m}{a}} \; . 
$\hfill} (8.2)

\

\h Here, because of the non-symmetric character of spectrum
(\ref{spectro}), let us call $\Delta\om_+$ ($>0$) the
bandwidth to the right, and $\Delta\om_-$ ($<0$) the bandwidth
to the left of $\om_\crm$; so that $\Delta\om = \Delta\om_+ -
\Delta\om_- $. It should be noted however that, already for
small values of $m$ (typically, for $m\geq 10$), one has
$\Delta\om_+ \approx - \Delta\om_-$. \ Once defined
$\om_\crm$ and $\Delta\om$, one can determine $m$ from the
first equation. Then, using the second one, $a$ is found.

\h Figure 3 illustrates the behavior of relation (8.1). From this
figure, one can observe that the smaller $\Delta\om / \om_\crm$
is, the higher $m$ must be. Thus, one can notice that $m$ plays
the fundamental role of controlling the spectrum bandwidth.

\h From the X-wave spectrum, it is known that $a$ is related to
the (negative) slope of the spectrum. \ Contrarily to $a$,
quantity $m$ has the effect of rising the spectrum. In this way,
one parameter compensates for the other, producing the
localization of the spectrum inside a certain frequency range. At
the same time, this fact also explains (because of relation
(8.2)) why an increase of both $m$ and $a$ is necessary to keep
the same $\om_\crm$. This can be seen from Fig.4.

\h In Fig.4, both spectra have {\em the same} $\om_\crm$. Taking
the narrow spectrum as a reference, one can observe that, to get
such a result, both quantities $m$ and $a$ have to increase.
Moreover, this figure shows the important role of $m$ for
generating a wider, or narrower, spectrum.

\

{\bf 3.2 -- X-type waves in a dispersionless medium }\\

 \h To illustrate the use of the proposed solutions, let us define the
ordinary X-wave, by rewriting eq.(\ref{solgeral}) with $S(\om)=
\exp[-a\om]$:

\

\begin{equation}
\begin{array}{l}\Psi (\rho,\ze) \ug \int_{0}^{\infty}J_0(\frac{\om}{V} \rho
\sqrt{n_0^2\frac{V^2}{c^2}-1}) \;
\dis{\e^{-(aV-i\ze) \om/V}} \drm\om \\
\\
\;\;\;\;\;\;\;\;\;\ug V\,\int_{0}^{\infty}J_0(\frac{\om}{V}
\rho \sqrt{n_0^2\frac{V^2}{c^2}-1}) \;
\dis{\e^{-(aV-i\ze) \om/V}} \drm(\frac{\om}{V}) \;\;\;\; ,
\end{array}
\label{intX}
\end{equation}

\

which is the same as
\

\begin{equation} X \; \equiv \; \Psi (\rho,\ze) \ug
\frac{V}{\sqrt{(aV-i\ze)^2+\rho^2(n_0^2\frac{V^2}{c^2}-1)}} \; , \;\;
\label{X}
\end{equation}

\

where $n_0$ is the refractive index of the medium underlying these
considerations. Applying our spectrum expressed by
eq.(\ref{spectro}), equation (\ref{solgeral}) can be rewritten as

\

\begin{equation}
\Psi (\rho,\ze)\ug
V\,\int_{0}^{\infty}(\frac{\om}{V})^n J_0 \left( \frac{\om}{V} \rho
\sqrt{n_0^2\frac{V^2}{c^2}-1} \right) \; \dis{\e^{-(aV-i\ze)\om/V}}
\drm(\frac{\om}{V})\,\,\,\,. \label{intXn}
\end{equation}

\

\h We have therefore seen that the use of a spectrum like (\ref{spectro}) allows
shifting it towards any frequency and confining it within the desired
frequency range. In fact, this is one of its most important characteristics.

\h It can be seen that eqs.(\ref{intX}) and (\ref{X}) are
equivalent. Deriving eq.(\ref{intX}) with respect to $(aV-i\ze)$, a
multiplicative factor $(\frac{\om}{V})$ is each time produced (an obvious
property of Laplace transforms). In this way, it is
possible to write eq.(\ref{intXn}) as :

\begin{equation}
\Psi (\rho,\ze)\ug (-1)^m\frac{\partial^m\,X}{\partial(aV-i\ze)^m}
\label{Xn} \ .
\end{equation}

\h A {\em different} expression for eq.(\ref{intXn}), without any need of
calculating the $m$-th derivative of the X-wave, can be found by using
identity (6.621) of ref.[14]:

\

$$
\;\;\;\;\;\;\;\;\;\;\;\;\;\;\;   \Psi (\rho,\ze)\ug
\frac{\Gamma(m+1)\,X^{m+1}}{V^{m}} \, \, F\left(\frac{m+1}{2}\, ,
\, -\frac{m}{2} \, ; \, 1 \, ; \,
(n_0^2\frac{V^2}{c^2}-1)\,\rho^2\, \frac{X^2}{V^2} \right) \; ,
\;\;\;\;\;\;\;\;\;\;(12')
$$

\

where $X$ is the ordinary X-wave given in eq.(\ref{X}), and $F$ is a
Gauss' hypergeometric function. Equation (12') can be useful in
the cases of large values of $m$.

\h Let us call attention to equations (\ref{Xn}) and (12'): to our
knowledge,\footnote{It can be noticed that $\partial X/\partial (aV-i\ze)=(iV)^{-1}
\partial X/\partial t$. Time derivatives of the X-wave have been actually
considered by J.Fagerholm et al.[15]: however the properties of
the spectrum generating those solutions (like its shifting in
frequency and its bandwidth) did not find room in that previous
work.} no analytical expression had been previously met for
X-{\em type} waves, which can be localized in the neighbourhood of any
chosen frequency with an adjustable bandwidth. Equations (\ref{Xn}) allow 
getting one or more of them in a simple way: All that has to be done is
calculating the $m$-th derivative of $X$ with respect to $(aV-i\ze)$. \
Alternatively, one can have recourse to eq.(12').

\h Fig.5 shows an example of an X-shaped wave for microwave
frequencies. To that aim, it was chosen $\om_\crm=6 \times 10^9 \;
\GHz$ and $\Delta\om=0.9 \, \om_\crm$, and the values of $n$
and $a$ were calculated by using eqs.(8.1) and (8.2): thus
obtaining $m=10$ and $a=1.6667\times 10^{-9}$. As one can see, the resulting
wave has really the same shape and the same properties as the
classical X-waves: namely, both a longitudinal and a transverse
localization.

\

\

{\large\bf 4. -- Superluminal localized waves in dispersive media}\\

\h We shall now pass to dealing with dispersive media.

\h In Section (2), equations (\ref{transform}) and (\ref{Krho})
were written for a dispersionless medium ($n_0=\constant$,
independent of the frequency). However, for a typical medium,
when the refractive index depends on the wave
frequency, $n(\om)$, those equations become[13]

\begin{equation}
\left\{
\begin{array}{l}
k_{\rho}(\om) = \frac{\om}{c} \, n(\om) \, \sin(\th) \\
\\
k_{z}(\om) = \frac{\om}{c} \, n(\om) \, \cos(\th) \; .
\end{array}
\right.  \label{transdisp}
\end{equation}

\h The above equations describe one of
the basic points of this work. In Section (2) it was
mentioned that $\th$ determines the wave velocity: a fact that can be
exploited when one looks for a localized wave that does not suffer
dispersion. In other words, one can choose a particular frequency
dependence of $\th$ to compensate for the (geometrical) dispersion
due to the variation with the frequency of the refractive index[13].

\h If the frequency dependence of the refractive index in a medium is known,
within a certain frequency range, let us see how the consequent dispersion
can be compensated for. When a dispersionless pulse is desired, the
constraint $k_{z} = a + \om \, b$ must be satisfied. And, by using the
last term in eq.(\ref{transdisp}), one infers that such a constraint is
forwarded by the following  relationship between $\th$ and $\om$:

\begin{equation}
\cos(\th(\om))=\frac{a + b\,\om}{\om\, n(\om)} \ ,
\label{thetaomega}
\end{equation}

where $a$ and $b$ are arbitrary constants (and $b$ is related to
the wave velocity: $b=1/V$). For convenience, we shall consider $a=0$.
Then, eq.(\ref{solgeral}) can be rewritten as

\begin{equation} \Psi (\rho,\ze) \ug
\int_{0}^{\infty}S(\om)\,J_0(\rho\om
\sqrt{\frac{n^2(\om)}{c^2}-b^2}\,) \; \dis{\e^{+i\om
b\ze}}\, \drm\om \; . \label{solgeraldisp}
\end{equation}

\h Let us stress that this equation is a priori suited for many
kinds of applications. In fact, whatever its frequency be (in the
optical, acoustic, microwave,... range), it constitutes the
integral formula representing a wave which propagates without
dispersion in a {\em dispersive} medium.

\h Now, let us mention how it is possible to realize relation
(\ref{thetaomega}) for optical frequencies. Although limited to
the case of the air, or of low-dispersion media, the {\em
axicon}[5,6,11,16] is one of the simplest means to realize it.
Another possibility is using ``spectral hole burning filters", or
holograms[17]. More in general, one can follow a procedure
similar to the one illustrated in Figure 6.

\h The process illustrated in Fig.6 is actually simple. In fact,
there is a different deviation of the wave vector for each
spectral component in passing through the chosen device (axicon,
hologram, and so on): and such a deviation, associated with the
dispersion due to the medium, makes the phase velocity equal for
each frequency. This corresponds to no dispersion for the
group-velocity. More details about the physics under
consideration can be found in Ref.[13].

\h Now, let us consider a nearly gaussian spectrum as that given
by eq.(\ref{spectro}), and assume the presence of a dispersive medium whose
refractive index (for the frequency range of interest) can be written in
the form

\begin{equation}
n(\om)=n_0 + \om \, \delta \; , \;\; \label{n}
\end{equation}

where $n_0$ is a constant, while $\delta$ is a free parameter that
makes it possible a linear behavior of $n(\om)$: something that
is actually realizable for frequencies far from the resonances
associated with the used material. Notice that the linear
relationship between the refractive index and the wave frequency
assumed in eq.(\ref{n}) is not necessary: but its existence
gets our calculations simplified.

\h In this way, substituting eq.(\ref{n}) into eq.(\ref{solgeraldisp}),
and considering the spectrum, shifted towards optical frequencies,
given by eq.(\ref{spectro}), a relation similar to eq.(\ref{intXdisp})
is found:

\begin{equation}
\Psi (\rho,\ze,\delta)\ug
V\,\int_{0}^{\infty}(\frac{\om}{V})^n J_0 \left(\frac{\om}{V} \rho
\sqrt{\frac{V^2}{c^2}(n_0+\delta\om)^2 - 1} \right) \;
\dis{\e^{-(aV-i\ze)\om/V}} \drm(\frac{\om}{V}) \; .
\label{intXdisp}
\end{equation}

\h To the purpose of evaluating eq.(\ref{intXdisp}), let us make a Taylor
expansion and rewrite it as

\begin{equation}
\Psi(\rho,\ze,\delta)\ug\Psi(\rho,\ze,0)\;+\;
\delta\;\;\frac{\partial\Psi}{\partial\delta} \left|_{\delta=0} \right. \;+\;
\frac{\delta^2}{2!}\;\;\frac{\partial^2\Psi}{\partial\delta^2} \left|_{\delta=0}
\right. \;+\; \frac{\delta^3}{3!}\;\;\frac{\partial^3\Psi}{\partial\delta^3}
\left|_{\delta=0} \right. \;+\; ...
\label{taylor}
\end{equation}

\h For the above equation it is known that, if $\delta$ is small
enough, it is possible to truncate the series at its first
derivative. For the time being, let us assume this is the case and
that there is no problem on truncating eq.(\ref{taylor}). One can
check Fig.7, which shows typical values of $\delta$ for
$\SiO_{2}$, a typical raw-material in fiber optics.

\h Looking at eq.(\ref{taylor}), one can notice that its first term
$\Psi (\rho,\ze,0)$ is already known to us, because it coincides with
the solution given by our eq.(\ref{intXn}). To complete the expansion
(\ref{taylor}), one must find $\frac{\partial\Psi}
{\partial\delta}|_{\delta=0}$. \ After some simple
mathematical manipulations, one gets that

\begin{equation} \frac{\partial\Psi}{\partial\delta}\,\left|_{\delta=0}\right.
\ug -\frac{V^4\,\rho\,n_0}{c^2\,\sqrt{n_0^2\frac{V^2}{c^2}-1}} \
\int_{0}^{\infty}(\frac{\om}{V})^{m+2} \ \e^{-(aV-i\ze)\om/V} \
J_1 \left( \rho\frac{\om}{V} \sqrt{n_0^2\frac{V^2}{c^2}-1}\, \right)
\; \drm(\frac{\om}{V}) \;. \label{deltapsi}
\end{equation}

\h This integral can be easily evaluated by using identity 6.621-4 of
ref.[14], so to obtain

\

\begin{equation} \frac{\partial\Psi}{\partial\delta}\,\left|_{\delta=0} \right.
\ug (-1)^{m+4} \; \frac{V^3\,n_0}{c^2\,\left(n_0^2\frac{V^2}{c^2}-1\right)} \
\frac{\partial^{m+2}}{\partial(aV-i\ze)^{m+2}}
\left[(aV-i\ze)\,X\right] \; , \label{deltapsi2}
\end{equation}

\

\h As in the case of eqs.(12), (12'), {\em another} form for
expressing eq.(\ref{deltapsi}) can be found by having recourse
once more to the identity (6.621) of ref.[14]:

\

$$
\;\;\;\;\;\;  \frac{\partial\Psi}{\partial\delta}\, \left|_{\delta=0} \right.
\ug -\frac{n_0\,\rho^2 \, \Gamma(m+4)\,X^{m+4}}{2\,c^2\,V^{m}} \, \,
F\left(\frac{m+4}{2}\, , \, -\frac{-m-1}{2} \, ; \, 2 \, ; \,
(n_0^2\frac{V^2}{c^2}-1)\,\rho^2\, \frac{X^2}{V^2} \right)
\;\;\;\;\;(20')
$$

\

where, as before, $X$ is the ordinary X-wave given by eq.(\ref{X}) and $F$ is
again a Gauss' hypergeometric function.  Once more, equation (20') can be
useful in the cases of large values of $m$.

\h Finally, from our basic solution (\ref{Xn}) and its first
derivative (\ref{deltapsi2}), one can write the desired
solution of eq.(\ref{solgeraldisp}) as

\begin{equation}
\Psi(\rho,\ze)\ug
(-1)^n\frac{\partial^n\,X}{\partial(aV-i\ze)^n}\,+\,
(-1)^{n+4}\frac{V^3\,n_0}{c^2\,\left(n_0^2\frac{V^2}{c^2}-1\right)} \
\frac{\partial^{n+2}}{\partial(aV-i\ze)^{n+2}}
\left[(aV-i\ze)\,X\right] \, \delta \; . \label{psitot}
\end{equation}

\h However, if one wants to use equations (12') and (20'),
instead of eqs.(\ref{Xn}) and (\ref{deltapsi2}), the  solution
(\ref{psitot}) can be written in the form

\

$$\begin{array}{l} \;\;\;\;\;\;\;\dis{\Psi(\rho,\ze)\ug
\frac{\Gamma(m+1)\,X^{m+1}}{V^{m}} \, \,
F\left(\frac{m+1}{2}\, , \, -\frac{m}{2} \, ; \, 1 \, ; \,
(n_0^2\frac{V^2}{c^2}-1)\,\rho^2\, \frac{X^2}{V^2} \right)} \\
\\
\;\;\;\;\;\;\;\;\;\;\;\;\;\;\;\;\;\;\;\;\;\;\;-\,\delta\,
\frac{n_0\,\rho^2\Gamma(m+4)\,X^{m+4}}{2\,c^2\,V^{m}}
\, \, F\left(\frac{m+4}{2}\, , \, -\frac{-m-1}{2} \, ; \, 2 \, ;
\, (n_0^2\frac{V^2}{c^2}-1)\,\rho^2\, \frac{X^2}{V^2}
\right) \ , \;\;\;\;\;\;\;\;\;\;\;\;\;  \end{array} (21')$$

\

\h It is also interesting to notice that, e.g., the approximated Superluminal
localized solution (\ref{psitot}) for a dispersive medium has been obtained
from simple mathematical operations (derivatives) applied to the standard
``X-wave''.

\

\

{\large\bf 5. -- Optical applications}\\

\h To illustrate what was said before, two practical examples will
be considered, both in optical frequencies. When mentioning
optics, it is natural to refer ourselves to optical fibers. Then,
let us suppose the bulk of the dispersive medium under
consideration to be fused Silica ($\SiO_2$).

\h Far from the medium resonances (which is our case), the
refractive index can be approximated by the well-known Sellmeier
equation[18]

\begin{equation}\label{selm}
n^2(\om)\ug\,1\,+\,\sum_{j=1}^{m}\,\frac{B_j\,\om_j^2}{\om_j^2-
\om^2} \, ,
\end{equation}

where $\om_j$ is the resonance frequency, $B_j$ is the strength
of the $j$th resonance, and $N$ is the total number of the
material resonances that appear in the frequency range of
interest. For typical frequencies of ``long-haul transmission" in
optics, it is necessary to choose $N=3$, which leads us to the
values[18] $B_1=0.6961663$, $B_2=0.4079426$, $B_3=0.8974794$,
$\lambda_1=0.0684043 \; \mu \mrm$, $\lambda_2=0.1162414 \; \mu
\mrm$ and $\lambda_3=9.896161 \; \mu \mrm$.

\h Figure 7 illustrates the relation between $N$ and $\om$,
and specifies the range that will be adopted here. In the two
examples, the spectra are localized around the angular frequency
$\om_\crm=23.56 \times 10^{14} \;$Hz (which corresponds to the
wavelength $\la_\crm=0.8\mu \mrm$), with two different bandwidth
$\Delta\om_1 = 0.55 \om_\crm$ and $\Delta\om_2 = 0.4 \om_\crm$. The
values of $a$ and $N$ corresponding to these two situations are
$a=1.14592 \times 10^{-14}$, $N=27$, and $a = 1.90986 \times
10^{-14}$, $N=41$, respectively.

\h Looking at these ``windows'', one can notice that Silica does
not suffer strong variations of its refractive index. As a matter
of fact, a linear approximation to $n=n(\om)$ is quite
satisfactory in these cases. Moreover, for both situations, and
for their respective $n_0$ values, the value of parameter
$\delta$ results to be very small, verifying condition
(\ref{solgeraldisp}): which means that it is quite acceptable our
truncation of the Taylor expansion. The beam intensity profiles
for both bandwidths are shown in Figs. 8 and 9.

\h In the first figure, one can see a pattern similar to that of
Fig.2; but here, of course, the pulse is much more localized
spatially and temporally (typically, it is a fentosecond pulse).

\h In the second figure, one can observe some little differences with
respect to the first one, mainly in the spatial oscillations inside the
wave {\em envelope\/}[19]. \  This may be explained by taking
into account that, for certain values of the bandwidth, the
carrier wavelength become shorter than the width of the spatial
envelope; so that one meets a well defined carrier frequency.

\h Let us point out that both these waves are transversally
{\em and} longitudinally localized, and that, since the dependence of
$\Psi$ on $z$ and $t$ is given by $\ze =z-Vt$, they are free from dispersion,
just like a classical X-shaped wave.

\

\

{\large\bf 6. -- Conclusions}\\

\h In this paper we have first worked out analytical Superluminal
localized solutions to the wave equation for arbitrary frequencies
and with adjustable bandwidth in vacum. The same methodology has been
then used to obtain new, analytical expressions representing
X-shaped waves (with arbitrary frequencies and adjustable
bandwidth) which propagate {\em in dispersive media}. Such expressions
have been
obtained, on one hand, by adopting the appropriate spectrum
(which made possible to us both choosing the carrier frequency rather
freely, and  controlling the spectral bandwidth), and,
on the other hand, by having recourse to simple mathematics.
Finally, we have illustrated some examples of our approach with
applications in optics, considering fused Silica as the
dispersive medium.

\

\

{\bf Acknowledgements} -- The authors are very grateful to Cesar
Dartona and Amr Shaarawi for continuous scientific collaboration;
and to Jane Marchi Madureira for stimulating discussions. \ For useful
discussions they thank also C.Becchi, M.Brambilla, C.Cocca, R.Collina,
G.C.Costa, P.Cotta-Ramusino, G.Degli Antoni, F.Fontana, M.Villa and
M.T.Vasconselos.

\newpage

{\large\bf Figure Captions}\hfill\break

Figure 1 -- Geometrical representation, in the plane ($\om,k_z$),
of the condition (2): see the text.\hfill\break

Figure 2 -- Illustration of the real part of an X-wave with
bandwidth, $\Delta\om$, of 10 MHz starting from zero.\hfill\break

Figure 3 -- Behavior of the derivative number, $m$, as a function of
the normalized bandwidth frequency, $\Delta\om_{\pm}/\om_\crm$.
Given a central frequency, $\om_\crm$, and a bandwidth,
$\Delta\om_{\pm}$, one finds the exact value of $m$ by substituting these
values into eq.(8.1).\hfill\break

Figure 4 -- Normalized spectra for $\om_\crm=23.56 \times 10^{14} \;$Hz
and different bandwidths. The first with $N=27$ (solid
line), and the second spectrum with $N=41$ (dotted line). See the text.\hfill\break

Figure 5 -- The real part of an X-shaped beam for microwave
frequencies in a dispersionless medium.\hfill\break

Figure 6 -- Sketch of a generic device (axicon,
hologram, etc.) suited to properly deviating the wave vector of each
spectral component.\hfill\break

Figure 7 -- Variation of the refractive index $n(\om)$ with
frequency for fused Silica. The solid line is its behaviour,
according to Sellmeir's formulae. The open circles and squares
are the linear approximations for $N=41$ and $N=27$,
respectively.\hfill\break

Figure 8 -- The real part of an X-shaped beam for optical
frequencies {\em in a dispersive medium}, with $N=27$. It refers to the 
larger window in Fig.7.\hfill\break

Figure 9 -- The real part of an X-shaped beam for optical
frequencies {\em in a dispersive medium}, with $N=41$.  It refers to the
inner window in Fig.7.\hfill\break

\newpage

\centerline{{\bf References}}

\

[1] J.Durnin, J.J.Miceli and J.H.Eberly: Phys. Rev. Lett. 58 (1987) 1499;
Opt. Lett. 13 (1988) 79; J.N.Brittingham: J. Appl. Phys. 54 (1983)
1179.\hfill\break

[2]  R.W.Ziolkowski: J. Math. Phys. 26 (1985) 861;  Phys. Rev.
A39 (1989)2005.\hfill\break

[3] A.Shaarawi, I.M.Besieris and R.W.Ziolkowski: J. Math. Phys. 30
(1989) 1254; \ A.Shaarawi, R.W.Ziolkowski and I.M.Besieris: J.
Math. Phys. 36 (1995) 5565.\hfill\break

[4] R.Donnelly and R.W.Ziolkowski: Proc. Roy. Soc. London A440
(1993) 541.\hfill\break

[5] J.-y.Lu and J.F.Greenleaf: IEEE Trans. Ultrason. Ferroelectr.
Freq.Control 39 (1992) 441. \ In this case the beam speed is
larger than the {\em sound} speed in the considered
medium.\hfill\break

[6] E.Recami: Physica A252 (1998) 586, and refs. therein.\hfill\break

[7] I.M.Besieris, M.Abdel-Rahman, A.Shaarawi and A.Chatzipetros:
Progress in Electromagnetic Research (PIER) 19 (1998)
1.\hfill\break

[8] M.Zamboni-Rached, E.Recami and H.E.Hern\'andez-Figueroa:
``New localized Superluminal solutions to the wave equations with
finite total energies and arbitrary frequencies", e-print \#
physics/0109062, in press in Europ. Phys. Journal-D.\hfill\break

[9] M.Zamboni-Rached: ``Localized solutions: Structure and
Applications", M.Sc. thesis; Phys. Dept., Campinas State
University, 1999).\hfill\break

[10] J.-y.Lu and J.F.Greenleaf: IEEE Trans. Ultrason.
Ferroelectr. Freq.Control 39 (1992) 19.\hfill\break

[11] P.Saari and K.Reivelt: ``Evidence of X-shaped
propagation-invariant localized light waves", Phys. Rev. Lett. 79
(1997) 4135.\hfill\break

[12] D.Mugnai, A.Ranfagni and R.Ruggeri: Phys. Rev. Lett. 84
(2000) 4830. \ For a panoramic view of the whole experimental situation, cf.
E.Recami: Found. Phys. 31 (2001) 1119 [e-print \# physics/0101108].\hfill\break

[13] P.Saari and H.S\~{o}najalg: Laser Phys. 7 (1997) 32.\hfill\break

[14] I.S.Gradshteyn and I.M.Ryzhik: {\em Integrals, Series and
Products}, 4th edition (Ac.Press; New York, 1965). \hfill\break

[15] J.Fagerholm, A.T.Friberg, J.Huttunen, D.P.Morgan and
M.M.Salomaa: Phys. Rev. E54 (1996) 4347.\hfill\break

[16] R.M.Herman and T.A.Wiggins: J. Opt. Soc. Am. A8 (1991)
932. \hfill\break

[17] Cf., e.g., H.S\~{o}najalg, A.Gorokhovskii, R.Kaarli, et al.: Opt. Commun. 71
(1989) 377. \hfill\break

[18] G.P.Agrawal: {\em Nonlinear Fiber Optics}, 2nd edition (Ac.Press;
New York, 1995). \hfill\break

[19] Cf., e.g., also M.Zamboni-Rached, K.Z.N\'obrega, E.Recami \&
H.E.Hern\'andez F.: ``Superluminal X-shaped beams propagating
without distortion along a co-axial guide", to appear in Phys. Rev. E.
\hfill\break

\end{document}